\documentclass[showpacs,preprintnumbers,amsmath,amssymb]{revtex4}

\usepackage{graphicx}
\usepackage{dcolumn}
\usepackage{bm}

\begin{document}

\preprint{APS/123-QED}

\title{Quantum Lithography in Macroscopic Observations}%

\author{De-Zhong Cao}
 \affiliation{Department of Physics, Applied Optics Beijing Area Major Laboratory, Beijing
Normal University, Beijing 100875, China.}%
\author{Kaige Wang}%
 \email{wangkg@bnu.edu.cn}
\affiliation{ CCAST (World Laboratory), P. O. Box 8730, Beijing
100080, China.\\
Department of Physics, Applied Optics Beijing Area Major
Laboratory, Beijing Normal University, Beijing 100875, China.}%

\date{\today}%

\begin{abstract}
We study the generalized Young's double-slit interference for the
beam produced in the spontaneously parametric down-conversion
(SPDC). We find that the sub-wavelength lithography can occur
macroscopically in both the two-photon intensity measurement and
the single-photon spatial intensity correlation measurement. We
show the visibility and the strength of the interference fringe
related to the SPDC interaction. It may provide a strong quantum
lithography with a moderate visibility in practical application.
\end{abstract}


\maketitle

Young's double-slit interference is a sign of waves, both the
classical wave and the quantum de Broglie wave for particles. For
a particle, the de Broglie wavelength depends on its mass. When
two particles with the same mass combine into a whole, for
example, a molecular, the corresponding de Broglie wavelength
reduces to the half of that for the single particle. Recently, a
similar effect for photons has been investigated both
experimentally and theoretically. \cite{joc}-\cite{nag} Due to the
fact of that this effect can overcome the Rayleigh diffraction
limit, it may have a prospective application in photon lithography
technology. Obviously, the wavelength reduction in Young's
interference is a pure quantum phenomenon, and it cannot occur for
any classical optical wave. The previous theoretical analysis
shows that the quantum lithography effect is based on the optical
field in a pure two-photon or multi-photon state. In the
experimental observation of the quantum lithography, the detection
of single photon pair is carried out by a beam splitter and the
two-photon coincidence detection. \cite{fon, da} These studies
show that the quantum lithography behaves in a microscopic realm,
which limits its practical application because of the weak
strength.

The recent investigations show that some quantum effects exhibited
in the microscopic observation can also occur in the macroscopic
realm. \cite{nag, gigi} In this paper, we show that the
sub-wavelength Young's interference can appear in two macroscopic
observations: the two-photon intensity measurement and the
single-photon spatial intensity-correlation measurement. The
source in the model is the spontaneously parametric
down-conversion (SPDC) of a type I crystal. We formulate the
second-order correlation function of the field
$G^{(2)}(x_{1},x_{2})=\left\langle E^{(-)}(x_{1})E^{(-)}(x_{2})E^{(+)}%
(x_{2})E^{(+)}(x_{1})\right\rangle $ in the interference plane. In
a weak interaction of SPDC when the two-photon state dominates the
down-converted beam, the second-order correlation function
$G^{(2)}(x_{1},x_{2})$ is proportional to the two-photon
coincidence probability. But for a strong interaction of SPDC,
$G^{(2)}(x_{1},x_{2})$ illustrates a macroscopic intensity
correlation for a large number of photons. We find that the
visibility of the sub-wavelength lithography will not be washed
out even in the very strong SPDC process.

The schematic setup of Young's interference is shown in Fig. 1. In
observation of two-photon Young's interference, we consider two
kinds of measurements: a two-photon detector positioned at $x$
which scans the two-photon intensity distribution and two
single-photon detectors positioned at $x_{1}$ and $x_{2}$ which
measure two-photon coincidence probability or the spatial
intensity correlation. We designate $a_{in}(x,t)$,
$a_{out}(x,t)$,$\ a_{1}(x,t)$ and $a_{2}(x,t)$ the slowly varying
field operators in the input plane $P_{in}$, the output plane
$P_{out}$ of the crystal, the double-slit plane $P_{1}$ and the
detection plane $P_{2}$, respectively. The double-slit function
$D(x)$ is
defined as%
\begin{equation}
D(x)=\left\{
\begin{array}
[c]{c}%
1\\
0
\end{array}%
\begin{array}
[c]{c}%
\quad\quad x\in\lbrack-\frac{d+b}{2},-\frac{d-b}{2}]\text{ and }[\frac{d-b}%
{2},\frac{d+b}{2}]\\
\text{others}%
\end{array}
\right.  , \label{1}%
\end{equation}
where $b$ and $d$ are the width of each slit and the interval
between two slits, respectively. By ignoring the thickness of the
double-slit, the field operator in plane $P_{1}$ is obtained as
\begin{equation}
a_{1}(x,t)=a_{out}(x,t)D(x)+a_{vac}(x,t)[1-D(x)], \label{2}%
\end{equation}
where the vacuum field $a_{vac}$ is introduced for maintaining the
bosonic commutation relation. (Note that $D^{2}(x)=D(x)$.) Since
the vacuum field has no contribution to the normal-order
correlation, it can be omitted in the calculation. In the paraxial
approximation, the transverse field in the detection plane $P_{2}$
is written as \cite{mandel}
\begin{equation}
a_{2}(x,t)=\frac{1}{2\pi}\iint\widetilde{a}_{1}(q,\Omega)\exp[-\mathrm{i}%
\frac{q^{2}z}{2k}+\mathrm{i}qx-\mathrm{i}\Omega t]dqd\Omega, \label{3}%
\end{equation}
where $k$ is the wavenumber of the beam and $z$ is the distance
between the double-slit and the\ detection plane.
$\widetilde{a}_{1}(q,\Omega)$ is the Fourier transform of
$a_{1}(x,t)$. Substituting Eq. ~(\ref{2}) into Eq. ~(\ref{3}) and
considering the integration $\int_{-\infty}^{\infty
}e^{-\mathrm{i}mq^{2}+\mathrm{i}xq}dq=(1-\mathrm{i})\sqrt{\pi/(2m)}%
e^{\mathrm{i}x^{2}/(4m)} $, we obtain%
\begin{equation}
a_{2}(x,t)=\frac{1-\mathrm{i}}{2\pi}\sqrt{\frac{k}{2z}}e^{\mathrm{i}%
\frac{kx^{2}}{2z}}\iint\widetilde{a}_{out}(q,\Omega)\mathfrak{D}%
(q,x)\exp[-\mathrm{i}\Omega t]dqd\Omega, \label{4}%
\end{equation}
where%
\begin{equation}
\mathfrak{D}(q,x)=\frac{1}{\sqrt{2\pi}}\int D(x^{\prime})\exp[\mathrm{i}%
\frac{k}{2z}x^{\prime2}+\mathrm{i}(q-\frac{kx}{z})x^{\prime}]dx^{\prime}.
\label{5}%
\end{equation}
In the far-field limit $z\gg kd^{2}$, the scale of the
interference pattern is much larger than that of the double-slit,
Eq. ~(\ref{5}) can be approximate to
the Fourier transform of the double-slit function $D(x)$%
\begin{equation}
\mathfrak{D}(q,x)\approx\widetilde{D}(\frac{kx}{z}-q), \label{6}%
\end{equation}
where
\begin{equation}
\widetilde{D}(q)=\frac{2b}{\sqrt{2\pi}}\operatorname{sinc}(qb/2)\cos(qd/2).
\label{6p}%
\end{equation}
Therefore, Eq.\ ~(\ref{4}) can be simplified as
\begin{equation}
a_{2}(x,t)=\frac{1-\mathrm{i}}{2\pi}\sqrt{\frac{k}{2z}}e^{\mathrm{i}%
\frac{kx^{2}}{2z}}%
{\displaystyle\iint}
\widetilde{a}_{out}(q,\Omega)\widetilde{D}(\frac{kx}{z}-q)\exp[-\mathrm{i}%
\Omega t]dqd\Omega. \label{7}%
\end{equation}

In the classical regime, we consider a stationary monochromic
plane wave in a
coherent state $|\alpha\rangle$, i.e. $\left\langle \widetilde{a}%
_{out}(q,\Omega)\right\rangle =\alpha\delta(q)\delta(\Omega)$,
which is normally incident upon the double-slit. In the far-field
limit, by using Eq. ~(\ref{7}), the first-order and the
second-order correlation functions for the field in the plane
$P_{2}$ are respectively obtained as
\begin{equation}
G^{(1)}(x,t)\propto\left\langle
a_{2}^{\dagger}(x,t)a_{2}(x,t)\right\rangle
=\frac{k}{4\pi^{2}z}|\alpha|^{2}\widetilde{D}^{2}(\frac{kx}{z}), \label{8}%
\end{equation}
and%
\begin{equation}
G^{(2)}(x_{1},x_{2},t)\propto\left\langle a_{2}^{\dagger}(x_{1},t)a_{2}%
^{\dagger}(x_{2},t)a_{2}(x_{2},t)a_{2}(x_{1},t)\right\rangle =\frac{k^{2}%
}{16\pi^{4}z^{2}}|\alpha|^{4}\widetilde{D}^{2}(\frac{kx_{1}}{z})\widetilde
{D}^{2}(\frac{kx_{2}}{z}). \label{9}%
\end{equation}
Equation ~(\ref{8}) gives a single-photon intensity distribution
describing a classical Young's interference pattern. However, Eq.
~(\ref{9}) illustrates a two-photon Young's interference pattern
for $x_{1}=x_{2}=x$, which should be measured by a two-photon
detector. These interference patterns have the same interval of
fringe-stripe $\lambda(z/d)$, showing the same as the classical
wave interference.

For an optical parametric amplification (OPA) of the nonlinear
crystal of type I,\ the relationship between the input and the
output field is given by \cite{r7, r8}
\begin{equation}
\widetilde{a}_{out}(q,\Omega)=U(q,\Omega)\widetilde{a}_{in}(q,\Omega
)+V(q,\Omega)\widetilde{a}_{in}^{\dagger}(-q,-\Omega). \label{10}%
\end{equation}
The coefficients $U(q,\Omega)$ and $V(q,\Omega)$ are written as
\begin{subequations}
\label{11}%
\begin{equation}
U(q,\Omega)=\Theta(q,\Omega)\left[  \cosh\Gamma(q,\Omega)+\frac{\mathrm{i}%
\delta(q,\Omega)}{2\Gamma(q,\Omega)}\sinh\Gamma(q,\Omega)\right],
\label{a11}%
\end{equation}%
\begin{equation}
V(q,\Omega)=\Theta(q,\Omega)\frac{g}{\Gamma(q,\Omega)}\sinh
\Gamma(q,\Omega), \label{11b}%
\end{equation}
\end{subequations}
where $\Gamma(q,\Omega)=\sqrt{g^{2}-\delta^{2}(q,\Omega)/4}$ and
$\Theta(q,\Omega)=\exp\{\mathrm{i}[(k_{z}(q,\Omega)-k)l_{c}-\delta
(q,\Omega)/2]\}$. The dimensionless mismatch function
$\delta(q,\Omega)$ is given by
\begin{equation}
\delta(q,\Omega)=\delta_{0}+\Omega^{2}/\Omega_{0}^{2}-q^{2}/q_{0}^{2}\text{,}
\label{12}%
\end{equation}
where $q_{0}=\sqrt{k/l_{c}}$ and $\Omega_{0}=1/(k_{\Omega}^{\prime\prime}%
l_{c})$ characterize the spatial-frequency and frequency
bandwidths, respectively. $g$ is the dimensionless coupling
strength of the nonlinear interaction and $\delta_{0}$ is the
phase matching parameter.

Now we consider the model of two-photon double-slit interference
for the SPDC interaction shown in Fig. 1. The quantum state of the
input field is assumed to be in the vacuum state so that the
output field of the crystal is of a SPDC field. Without applying
the far field limit, we substitute Eqs. ~(\ref{10}) into Eq.
~(\ref{4}), and calculate the second-order correlation function of
the
field $a_{2}(x,t)$ in the detection-plane $P_{2}$%
\begin{equation}
G^{(2)}(x_{1},x_{2})\propto
M(x_{1},x_{1})M(x_{2},x_{2})+\left\vert M(x_{1},x_{2})\right\vert
^{2}+\left\vert N(x_{1},x_{2})\right\vert ^{2},
\label{13}%
\end{equation}
where
\begin{subequations}
\label{14}%
\begin{equation}
M(x_{m},x_{n})=\frac{k}{(2\pi)^{2}z}\iint\left\vert
V(q,\Omega)\right\vert
^{2}\mathfrak{D}(q,x_{m})\mathfrak{D}^{\ast}(q,x_{n})dqd\Omega,\text{\qquad
}(m,n=1,2) \label{14a}%
\end{equation}%
\begin{equation}
N(x_{m},x_{n})=\frac{k}{(2\pi)^{2}z}\iint V(q,\Omega)U(-q,-\Omega
)\mathfrak{D}(q,x_{m})\mathfrak{D}(-q,x_{n})dqd\Omega. \label{14b}%
\end{equation}
In order to obtain the analytical form, we consider the broadband
limit, $q_{0}\gg1/d$. As shown in Ref. \cite{wang}, $q_{0}$
characterizes the spatial-frequency bandwidth of the coefficients
$U(q,\Omega)$ and $V(q,\Omega)$. Since the scale of the
double-slit function $D(x)$ is described by the interval $d$, the
bandwidth of function $\mathfrak{D}(q,x)$ is characterized by
$1/d$. The broadband limit can be realized by a thin crystal. In
this limit, functions $U(-q,-\Omega)$ and $V(q,\Omega)$ are
approximately independent of $q$ in the range where function
$\mathfrak{D}(q,x)$ has significant value and hence are taken as
$U(0,-\Omega)$ and $V(0,\Omega)$, respectively. Therefore, we
obtain
\end{subequations}
\begin{subequations}
\label{15}%
\begin{align}
M(x_{m},x_{n})  &  \approx\frac{k}{(2\pi)^{2}z}\iint\left\vert
V(0,\Omega
)\right\vert ^{2}\mathfrak{D}(q,x_{m})\mathfrak{D}^{\ast}(q,x_{n}%
)dqd\Omega\label{a15}\\
&  =\frac{k}{(2\pi)^{3}z}%
{\displaystyle\iiiint}
\left\vert V(0,\Omega)\right\vert
^{2}D(x^{\prime})D(x^{\prime\prime
})e^{\mathrm{i}\frac{k}{2z}(x^{\prime2}-x^{\prime\prime2})-\mathrm{i}\frac
{k}{z}(x_{m}x^{\prime}-x_{n}x^{\prime\prime})+\mathrm{i}q(x^{\prime}%
-x^{\prime\prime})}dqd\Omega dx^{\prime}dx^{\prime\prime}\nonumber\\
&  =\frac{k}{(2\pi)^{2}z}\iint\left\vert V(0,\Omega)\right\vert ^{2}%
D^{2}(x^{\prime})e^{-\mathrm{i}\frac{k}{z}(x_{m}-x_{n})x^{\prime}}d\Omega
dx^{\prime}\nonumber\\
&  =f_{1}\cdot\widetilde{D}[\frac{k}{z}(x_{m}-x_{n})],\nonumber
\end{align}
and similarly,%
\begin{equation}
N(x_{m},x_{n})\approx f_{2}\cdot\widetilde{D}[\frac{k}{z}%
(x_{m}+x_{n})], \label{b15}%
\end{equation}
where $f_{1}=\frac{k}{(2\pi)^{3/2}z}\int\left\vert
V(0,\Omega)\right\vert ^{2}d\Omega$ and
$f_{2}=\frac{k}{(2\pi)^{3/2}z}\int V(0,\Omega)U(0,-\Omega
)d\Omega$. In Eq. ~(\ref{15}), $D^{2}(x)=D(x)$ is taken into
account. In result, Eq. ~(\ref{13}) has the analytical form
\end{subequations}
\begin{equation}
G^{(2)}(x_{1},x_{2})\propto|f_{2}|^{2}\left\{  \xi\cdot\widetilde{D}%
^{2}(0)+\xi\cdot\widetilde{D}^{2}[\frac{k}{z}(x_{1}-x_{2})]+\widetilde{D}%
^{2}[\frac{k}{z}(x_{1}+x_{2})]\right\}  , \label{16}%
\end{equation}
where $\xi=\left\vert f_{1}/f_{2}\right\vert ^{2}$. Figure 2 shows
a 3D plot of $G^{(2)}(x_{1},x_{2})$ in the broadband limit.

To show the quantum feature of the second-order correlation of the
field in this model, we discuss two kinds of measurements. One of
the measurements is so called the two-photon intensity detection
in which the detector reacts to a
two-photon absorption. When the two-photon detector scans in the plane $P_{2}%
$, one obtains the distribution $G^{(2)}(x,x)$, as shown on the
diagonal line $x_{1}=x_{2}$ of Fig. 2. The other one of the
measurements is the spatial intensity correlation, in which two
single-photon detectors are placed at positions $x_{1}$ and
$x_{2}$. For the continuous modes, the field operators at two
different positions can commutate each other, \cite{r7} so that
the intensity correlation of two positions is equal to
$G^{(2)}(x_{1},x_{2})$, i.e. $\left\langle
I(x_{1})I(x_{2})\right\rangle \propto\left\langle
a^{\dagger}(x_{1})a^{\dagger}(x_{2})a(x_{2})a(x_{1})\right\rangle
$. For instance, we set two single-photon detectors at a pair of
symmetric positions $x_{1}=-x_{2}$ and it measures $G^{(2)}(x,-x)$
which is shown on the another diagonal line $x_{1}=-x_{2}$ of Fig.
2. According to Eq. ~(\ref{16}), we obtain
\begin{subequations}
\label{17}%
\begin{equation}
G^{(2)}(x,x)\propto|f_{2}|^{2}\left\{  2\xi\cdot\widetilde{D}^{2}%
(0)+\widetilde{D}^{2}[\frac{k}{z}2x]\right\}  , \label{17a}%
\end{equation}%
\begin{equation}
G^{(2)}(x,-x)\propto|f_{2}|^{2}\left\{ (\xi+1)\cdot\widetilde
{D}^{2}(0)+\xi\cdot\widetilde{D}^{2}[\frac{k}{z}2x]\right\}  . \label{17b}%
\end{equation}
In comparison with Eqs. ~(\ref{8}) and ~(\ref{9}), Eqs.
~(\ref{17}) show the sub-wavelength interference pattern with a
fringe-stripe interval $(\lambda/2)(z/d)$ for these two
measurements. In the previous theoretical treatment, the quantum
lithography is explained by the pure two-photon state. However, we
show that the\ quantum lithograph can occur in a general SPDC
interaction in which the spontaneous beam contains a large number
of photons. Therefore, the measurements of the second-order
correlation function can be considered as the macroscopic
observations.

The visibilities of the fringes are evaluated as
$\widehat{V}_{1}=\frac {1}{1+4\xi}$ and
$\widehat{V}_{2}=\frac{1}{3+2/\xi}$ for $G^{(2)}(x,x)$ and
$G^{(2)}(x,-x)$, respectively. Figure 3 shows the visibilities as
functions of the parameter $\xi$. When $\xi<<1$, the quantum
lithography exists only for the two-photon intensity detection. As
$\xi$ increases, $\widehat{V}_{1} $ decreases and
$\widehat{V}_{2}$ increases monotonously. In the case $\xi=1 $,
two kinds of quantum lithography have the same visibility 20\%.
However, it
can be seen that $\widehat{V}_{1}>\widehat{V}_{2}$ ($\widehat{V}_{1}%
<\widehat{V}_{2}$) when $\xi<1$ ($\xi>1$).

In the practical application of optical lithography, not only the
visibility but also the strength of the stripe are important.
Nevertheless, the DC background of the interference pattern can be
overcome, for example, by coating a covered film on the base.
According to Eqs. ~(\ref{17}), the strengths of two sub-wavelength
fringes described by $G^{(2)}(x,x)$ and $G^{(2)}(x,-x)$ are
proportional to $|f_{2}|^{2}$ and $|f_{1}|^{2}$, respectively. In
Fig. 4, we plot $|f_{i}|^{2}$ ($i=1,2$) and $\xi$ as functions of
the coupling strength $g$ for the different phase matching
$\delta_{0}$. For a weak SPDC process, $g\ll1$, $\xi$ and
$|f_{2}|^{2}$ are also small. It shows a weak sub-wavelength
fringe with the better visibility only for the two-photon
intensity detection. This corresponds to the case in which the
pure two-photon state dominates the SPDC beam. As $g$ is
increased, $\xi$ and $|f_{i}|^{2}$ are increased, too.
Consequently, the visibility $\widehat{V}_{1}$ diminishes but
$\widehat{V}_{2}$ enhances. However, for the very strong SPDC
interaction, $g\gg1$, $\xi$ tends toward about 1 and $|f_{i}|^{2}$
keeps exponentially increasing. According to Fig. 3, two
visibilities reach 20\% while the strength of the sub-wavelength
lithograph can be enhanced greatly. Both Figs. 3 and 4 are helpful
to choose the operation of crystal for the optimum lithography
application. As some examples, by setting different phase matching
parameters $\delta_{0}$, we plot several sub-wavelength
interference patterns in Figs. 5, in which Figs. 5a-5c and Figs.
5d-5f show $G^{(2)}(x,x)$ for the two-photon intensity detection
and $G^{(2)}(x,-x)$ for spatial intensity-correlation detection,
respectively. The coupling strength $g=1$ given in Fig. 5
indicates the maximum amplification factor $\exp(2g)\approx7.4$ in
the optical parametric amplifier. \cite{r8}

In summary, we formulate the second-order correlation in a
double-slit interference for the SPDC beam in the broadband limit.
We show that the sub-wavelength interference can occur in a
general SPDC process, both the weak and the strong interactions.
The quantum lithography can be considered as macroscopic
observation since it can occur for a general SPDC process. The two
kinds of observations, the two-photon intensity detection and the
single-photon spatial intensity correlation detection, provide
alternative method. Moreover, the intensive interference beam
makes quantum lithography technology in practicability.
\end{subequations}
\begin{acknowledgments}
This research is founded by the National Fundamental Research
Program of China with No. 2001CB309310, and the National Natural
Science Foundation of China, Project Nos. 60278021 and 10074008.
\end{acknowledgments}

captions for figures

Figure 1 Schematic setup of the Young's double-slit interference
for the SPDC process.

Figure 2 3-D plot of the second-order correlation function $G(X_{1},X_{2}%
)$\ in which $X_{i}=\frac{kb}{2\pi z}x_{i}$ ($i=1,2$) is the
normalized position. We set $\delta_{0}=0$, $g=1.84$, and $d=5b$.

Figure 3 Visibilities as functions of the parameter $\xi$ for two
kinds of measurements.

Figure 4 Parameters $\xi$ and $|f_{2}|^{2}$ as functions of the
coupling strength $g$ for different phase matchings $\delta_{0}$.

Figure 5 Sub-wavelength interference fringes: (a)-(c) for
$G^{(2)}(x,x)$ and
(d)-(f) $G^{(2)}(x,-x)$.%

%

\begin{center}
\includegraphics[
height=3.0173in, width=2.0593in, angle=-90.0
]%
{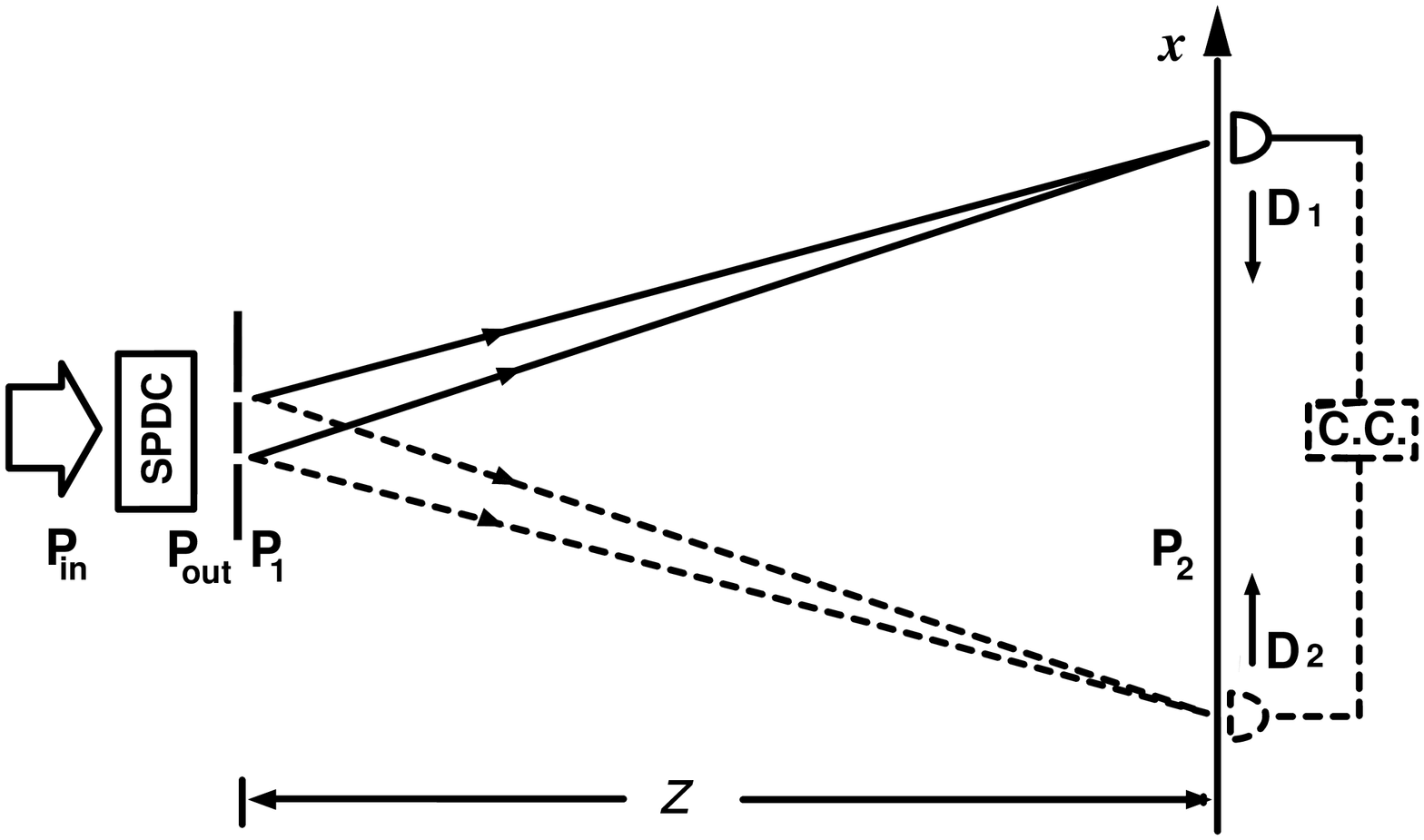}%
\\
Figure 1 Schematic setup of the Young's double-slit interference
for the SPDC process.
\end{center}

%

\begin{center}
\includegraphics[
height=3.0173in, width=2.0593in, angle=-90.0
]%
{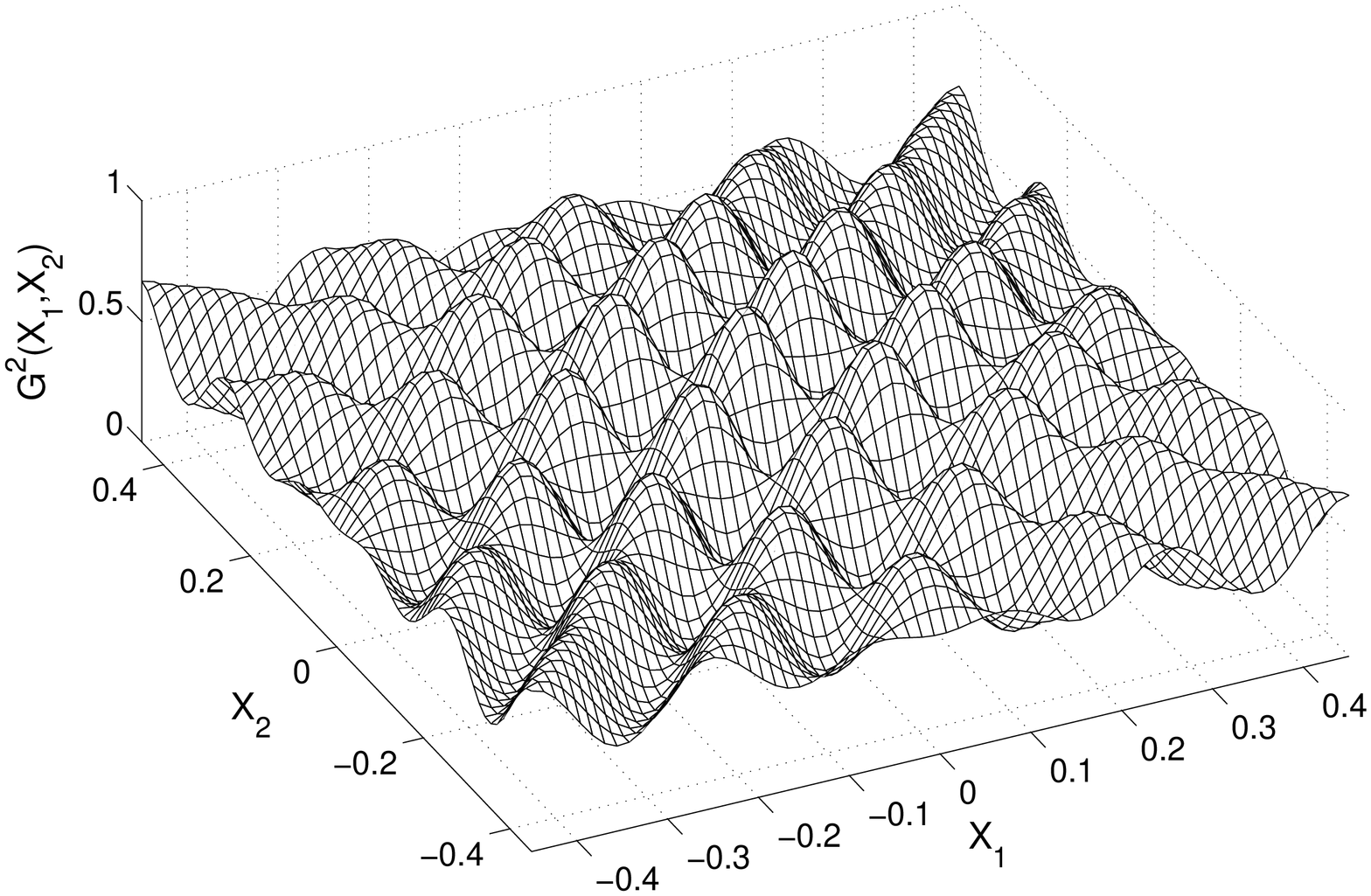}%
\\
Figure 2 3-D plot of the second-order correlation function $G(X_{1},X_{2}%
)$\ in which $X_{i}=\frac{kb}{2\pi z}x_{i}$ ($i=1,2$) is the
normalized position. We set $\delta_{0}=0$, $g=1.84$, and $d=5b$.
\end{center}

%

\begin{center}
\includegraphics[
height=3.2in, width=2.4in, angle=-90.0
]%
{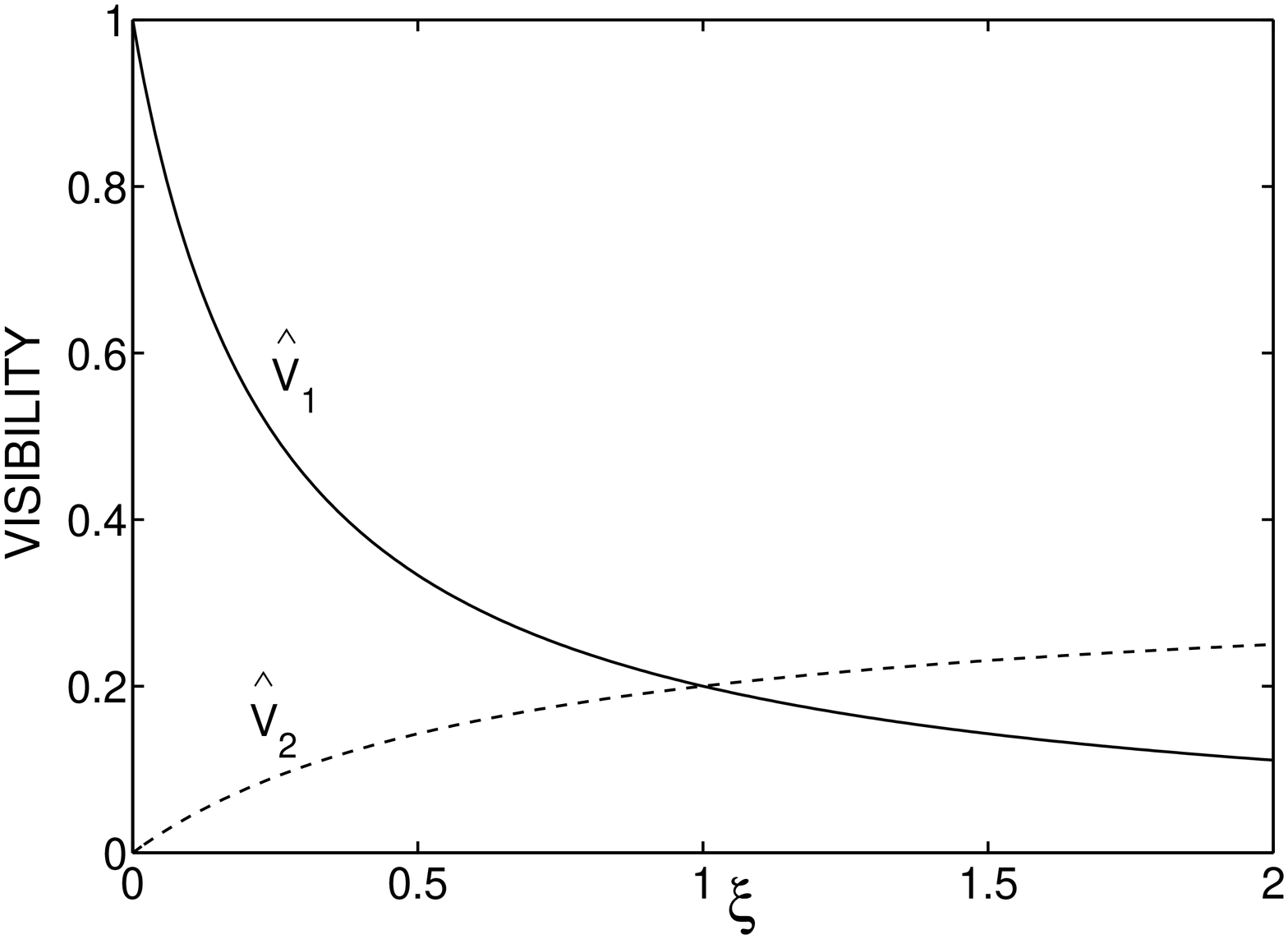}%
\\
Figure 3 Visibilities as functions of the parameter $\xi$ for two
kinds of measurements.
\end{center}
%

\begin{center}
\includegraphics[
trim=0.000000in 0.000000in 0.000000in 1.064131in, height=9.3457cm,
width=7.3284cm
]%
{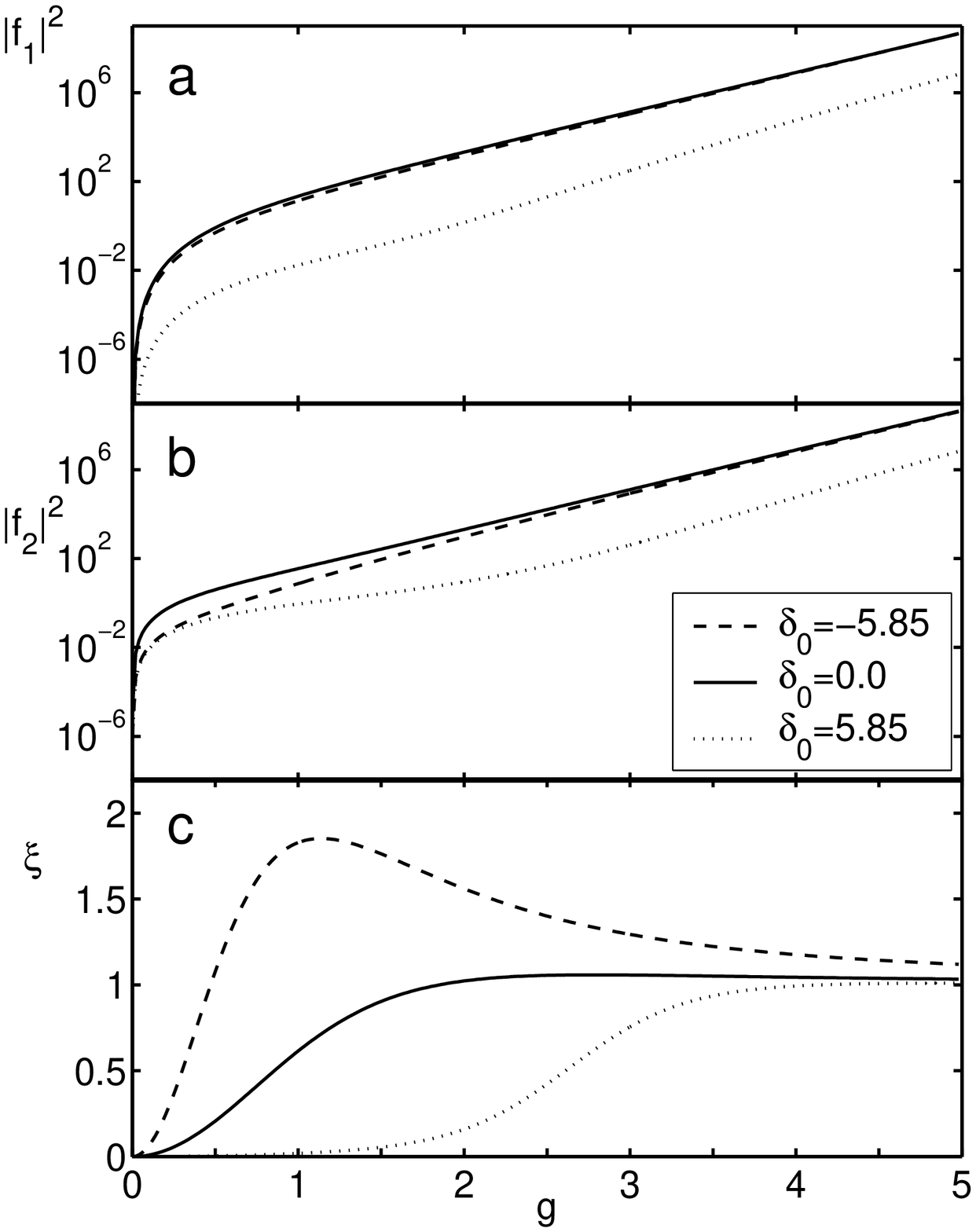}%
\\
Figure 4 Parameters $\xi$ and $|f_{2}|^{2}$ as functions of the
coupling strength $g$ for different phase matchings $\delta_{0}$.
\end{center}
%

\begin{center}
\includegraphics[
height=3.4564in, width=2.9281in
]%
{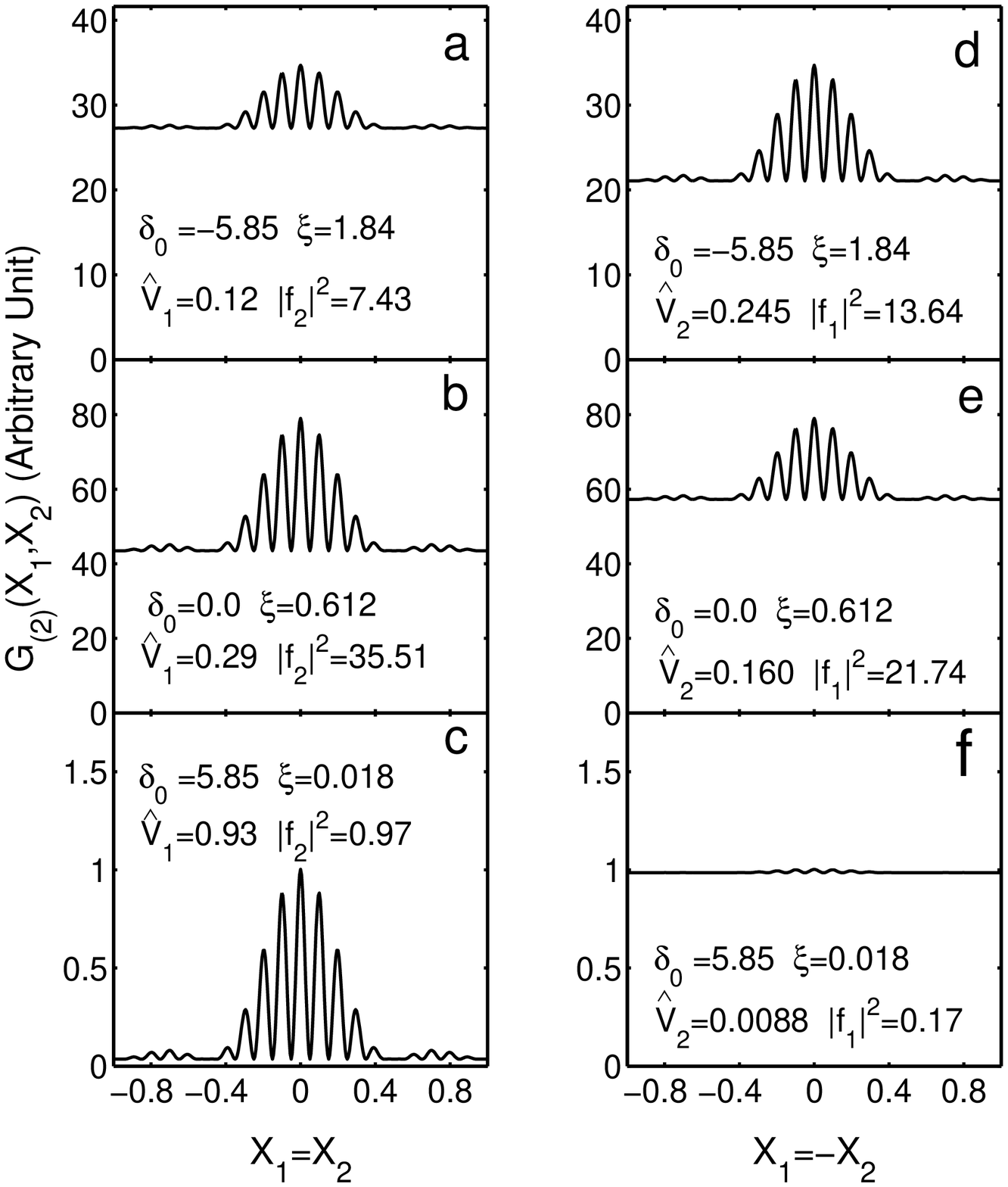}%
\\
Figure 5 Sub-wavelength interference fringes: (a)-(c) for
$G^{(2)}(x,x)$ and (d)-(f) $G^{(2)}(x,-x)$.
\end{center}

\end{document}